\documentclass[acmsmall,screen,authorversion,nonacm, table]{acmart}

\usepackage{microtype}
\usepackage{paralist}
\usepackage{array}
\usepackage{subfig}
\usepackage{graphicx}
\usepackage{multirow,booktabs}
\usepackage{graphicx}
\usepackage{marvosym}
\usepackage{ctable}
\usepackage{paralist}
\usepackage{hyperref}
\usepackage{verbatim}
\hypersetup{
    colorlinks=true,
    linkcolor=blue,
    citecolor=blue,      
    urlcolor=blue
}
\usepackage{makecell}
\usepackage{listings}
\usepackage{color}
\usepackage{adjustbox}
\usepackage{xcolor}
\usepackage{comment}
\usepackage{fancyhdr}
\usepackage{tcolorbox}
\usepackage{textpos}

\AtBeginDocument{%
  \providecommand\BibTeX{{%
    \normalfont B\kern-0.5em{\scshape i\kern-0.25em b}\kern-0.8em\TeX}}}
    
\setcopyright{acmcopyright}
\copyrightyear{2021}
\acmYear{2021}
\acmDOI{10.1145/1122445.1122456}

\acmJournal{JDIQ}
\acmVolume{37}
\acmNumber{4}
\acmArticle{111}
\acmMonth{8}

\begin{document}
\begin{textblock*}{140mm}(0.1mm,-2cm)
\begin{tcolorbox}[colframe=red!50, colback=yellow!50]
\footnotesize
This is the authors' preprint version. The article is accepted for publication in the Special Issue on
Data Transparency in the ACM Journal of Data and Information Quality (JDIQ). \copyright 2021

\end{tcolorbox}
\end{textblock*}


\title{Knowledge-driven Data Ecosystems Towards Data Transparency} 

\author{Sandra Geisler}
\email{sandra.geisler@fit.fraunhofer.de}
\affiliation{%
  \institution{Fraunhofer FIT and RWTH Aachen University, Germany}
  \streetaddress{Schloss Birlinghoven}
  \city{Sankt Augustin}
  \postcode{53757}
}
\author{Maria-Esther Vidal}
\email{maria.vidal@tib.eu}
\affiliation{%
  \institution{TIB-Leibniz Information Centre for Science and Technology, Gerrmany}
  \streetaddress{Welfengarten 1B}
  \city{Hannover}
  \postcode{30167}
}
\author{Cinzia Cappiello}
\email{cinzia.cappiello@polimi.it}
\affiliation{%
  \institution{Politecnico di Milano, Italy}
  \streetaddress{piazza Leonardo da Vinci 32}
  \city{Milano}
  \postcode{20133}
}
\author{Bernadette Farias L{\'o}scio}
\email{bfl@cin.ufpe.br}
\affiliation{%
  \institution{Federal University of Pernambuco, Brazil}
  \streetaddress{Cidade Universitaria}
  \city{Recife/PE}
  \postcode{50740-560}
}
\author{Avigdor Gal}
\email{avigal@ie.technion.ac.il}
\affiliation{%
  \institution{Technion Israel Institute of Technology, Israel}
  \streetaddress{Technion City}
  \city{Haifa}
  \postcode{32000}
}
\author{Matthias Jarke}
\email{jarke@dbis.rwth-aachen.de}
\affiliation{%
  \institution{RWTH Aachen University and Fraunhofer FIT, Germany}
  \streetaddress{Ahornstrasse 55}
  \city{Aachen}
  \postcode{52056}
}
\author{Maurizio Lenzerini}
\email{ lenzerini@diag.uniroma1.it}
\affiliation{%
  \institution{Sapienza Universit{\`a} di Roma, Italy}
  \streetaddress{via Ariosto 25}
  \city{Roma}
  \postcode{I-00185}
}
\author{Paolo Missier}
\email{paolo.missier@ncl.ac.uk}
\affiliation{%
  \institution{Newcastle University, United Kingdom}
  \streetaddress{Firebrick Avenue}
  \city{Newcastle upon Tyne}
  \postcode{NE4 5TG}
}
\author{Boris Otto}
\email{boris.otto@cs.tu-dortmund.de}
\affiliation{%
  \institution{TU Dortmund University, Germany}
  \streetaddress{Otto-Hahn-Str. 12}
  \city{Dortmund}
  \postcode{44227}
}
\affiliation{%
  \institution{Fraunhofer ISST, Germany}
  \streetaddress{Emil-Figge-Stra{\ss}e 91}
  \city{Dortmund}
  \postcode{44227}
}
\author{Elda Paja}
\email{elpa@itu.dk}
\affiliation{%
  \institution{IT University of Copenhagen, Denmark}
  \streetaddress{Rued Langgaards Vej 7}
  \city{Copenhagen S}
  \postcode{DK-2300}
}
\author{Barbara Pernici}
\email{barbara.pernici@polimi.it}
\affiliation{%
  \institution{Politecnico di Milano, Italy}
  \streetaddress{piazza Leonardo da Vinci 32}
  \city{Milano}
  \postcode{20133}
}
\author{Jakob Rehof}
\email{jakob.rehof@cs.tu-dortmund.de}
\affiliation{%
  \institution{TU Dortmund University, Germany}
  \streetaddress{Otto-Hahn-Str. 12}
  \city{Dortmund}
  \postcode{44227}
}
\affiliation{%
  \institution{Fraunhofer ISST, Germany}
  \streetaddress{Emil-Figge-Stra{\ss}e 91}
  \city{Dortmund}
  \postcode{44227}
}

\renewcommand{\shortauthors}{S. Geisler, M.-E. Vidal  et al.}


\begin{abstract}
A \emph{Data Ecosystem} offers a keystone-player or alliance-driven
infrastructure that enables the interaction of different stakeholders and the
resolution of interoperability issues among shared data. However, 
despite years of research in data governance and management, trustability is
still affected by the absence of transparent and traceable data-driven
pipelines. In this work, we focus on requirements and challenges that data
ecosystems face when ensuring data transparency. Requirements are derived from
the data and organizational management, as well as from broader legal and
ethical considerations. We propose a novel knowledge-driven data ecosystem
architecture, providing the pillars for satisfying the analyzed requirements. We
illustrate the potential of our proposal in a real-world scenario. Lastly, we
discuss and rate the potential of the proposed architecture in the fulfillment
of these requirements.
\end{abstract}

\maketitle
\section{Introduction}
\label{sec:introduction}

Industrial digitalization and the use of information technologies in public and
private sectors provide evidence of the pivotal role of data.
However, despite the paramount relevance of data-driven technologies,
organizations demand alliance-driven infrastructures capable of supporting
controlled data exchange across diverse stakeholders and transparent data
management.

\emph{Data Ecosystems (DEs)} are distributed, open, and adaptive information
systems with the characteristics of being self-organizing, scalable, and
sustainable \cite{OliveiraL18}. While centered on data, the main concern with
DEs is about knowledge generation and sharing. Thus, they aim to solve issues
like learning from unstructured and heterogeneous data, and construct new
abstractions and mappings. They may also offer various data-centric services,
including query processing and data analytics.

DEs are equipped with computational methods to exchange and integrate data while
preserving personal data privacy, data security, and organizational data
sovereignty.
The report of the Dagstuhl Seminar 19391 (September 22-27,
2019)\footnote{\url{http://www.dagstuhl.de/19391}} on {\em Data Ecosystems:
Sovereign Data Exchange among Organizations}  \cite{capiello_et_al:DR:2020:11845} contains
summaries of discussions and abstracts of talks from the seminar on various
topics, including requirements, use cases, and architectures.
Diverse reference architectures rely on DE foundations \cite{BaderML19,
LVST2020}. Keystone player-driven data ecosystems and B2C platforms like Google,
Alibaba, or Facebook are hugely successful. In contrast, the adoption of
alliance-driven platforms which aim at more equitable control and data sharing
\cite{OJ2019} is still lagging, even in crucial domains such as data-driven B2B
engineering collaboration \cite{jarke2020} or biomedicine. This paper focuses on
the alliance-driven setting, even though many addressed issues occur also in the
other category.

A few works address general data quality (DQ) aspects of DEs (e.g.,
\cite{DBLP:series/dcsa/BatiniS16,Kitsios2017,Donker2017,DBLP:journals/bise/ZhangIS19}).
In \cite{Donker2017}, the authors focus on open DEs and claim that data
availability and DQ need to be guaranteed, so as to prevent users to be hesitant
to use data. Kitsios et al. \cite{Kitsios2017} depict DQ assessment as one of
the fundamental components for building and maintaining a DE. DQ assessment
requires the definition of a DQ model composed of DQ dimensions and metrics.
Several DQ dimensions have been defined in the literature, as discussed in
\cite{DBLP:series/dcsa/BatiniS16}. Many dimensions have a possible impact on
data fairness and trustability, in particular completeness, accuracy, and
consistency, which have a significant impact both, on transparently processing
data for analysis and on data pipelines.
The lack of accountability for data transparency is one of the severe
limitations of existing interoperable methods and represents a critical aspect
of data quality. This paper starts from the hypothesis that these limitations
could be a significant reason for the slow adoption of DEs.

To account for data transparency, the rest of this paper offers the following
contributions: (1) an analysis of the specific requirements arising for DEs and
from DEs regarding data governance and transparency aspects; (2) a new form of
\emph{networks of knowledge-driven DEs} towards trustworthiness and wider
adoption; (3) challenges that stem from the identified requirements; and (4) an
assessment of how the various DE types address these requirements.

\section{Requirements of Transparent Data Ecosystems Motivated by an Example}
\label{motivation}
We motivate the need for expressive data ecosystems with an example from the health domain. 
Subsequently, we grasp the requirements demanded for data transparency in similar scenarios. 
 
\subsection{Motivating Example}
Consider a use case of multi-site clinical studies as an example to illustrate
the impact that managing multiple stakeholders have on interoperability and
transparency requirements.
In these studies, several parties are involved; they include clinics, resident
doctors, data scientists, patients, study nurses, quality assurance,
researchers, and care services.
A stakeholder may have multiple sources generating data.
For example, clinicians conduct examinations and collect, amongst others, sensor
readings, medical images, test results, and diagnostic reports.
These data collections are processed (e.g., transformed, curated, and
integrated); for transparency reasons, they are potentially annotated with
meta-data, domain vocabularies, and data quality values.

Data is analyzed to uncover insights that can support clinicians to conduct
thoughtful diagnostics and effective treatments.
Data management tasks are also influenced by the organization's regulations or
higher instances, such as regulations for data protection or rules defined by
the hospital, and strategic decisions. Patients may require transparency about
both, their treatment and the privacy protection of their data in cross-clinical
studies.
Each of these data management tasks brings up already multiple challenges for
data transparency and data quality management.
Additionally, data is exchanged between stakeholders to fulfill the goals of the
studies; data collected from the various sites has to be pseudonymized and
integrated to be audited by quality assurance. But transparency of these
processes must be maintained to protect against scientific fraud.
Further integration with data from additional parties, such as health insurance
companies, may be needed to be finally analyzed by study researchers.

The study setting corresponds to a \emph{network of knowledge-driven DEs}. This
network aligns the stakeholder DEs and their data; it also uses meta-data to
describe the network and its constituents. Furthermore, the network is
influenced by regulations, contracts, or agreements specific to the study at
hand. They may include participation agreements created by insurance companies
and patients consent forms authorizing data usage for specific studies.
Heterogeneity issues across the different network DEs impose challenges for DQ
management.
Moreover, documenting computational methods performed to assess and curate data
quality issues is crucial to guarantee data transparency.

\subsection{Requirements Analysis}
\label{sec:requirements}
The motivating example highlights the multiple issues that a DE needs to cover
in order to enhance trustability of the involved stakeholders. These issues are
not only present in biomedical applications, but rather exist in any application
where crucial decisions are driven by data \cite{bdva2020}.
Based on literature and reports from current European data sharing and data
space projects \cite{bdva2020,otto2019,curry2018}, requirements can be
categorized along data management, organizational aspects, and legal and ethical
issues.
In terms of \emph{data management}, tackling the challenges outlined in the
motivating example, demands (meta)data sharing among different stakeholders in a
secured and traceable manner. At an \emph{organization level}, trustable data
exchange requires complex ecosystems that underlie organizational-specific
business models, processes, and strategies to enforce sovereignty, privacy, and
protection of both, data and analytical outcomes.
Furthermore, sharing sensitive and personal data, e.g., clinical records, should
comply with data protection regulations and legal compliance at national and
international levels.
More importantly, accounting for ethical decisions made by stakeholders and
algorithms is crucial and the ability to provide reliable and verifiable
explanations of these decisions. Meeting these requirements at a \emph{legal and
ethical level} empowers DEs to safeguard data privacy and mitigate unfairness in
data-driven pipelines. Moreover, the satisfaction of these requirements provides
the foundations for ensuring that clinical data is only used according to
consents given by these data owners, i.e., the patients.
Next, each of these three requirement categories is described in more detail.

\paragraph{Data management requirements}
DEs, as described in the motivating example, demand sharing of data with a high
variety (e.g., in terms of type, structure, size, or frequency).
The requirements listed in this category concern both, data and metadata.
Data quality management has to be able to \textbf{(DMR1)} \emph{handle all kinds
of data and offer common DQ tools} to describe, query, and assess quality values
for the data.
In the medical domain for example, this comprises unstructured data, such as
images and texts, but also highly structured data from databases, csv files, and
data streams.
Additionally, \textbf{(DMR2)} \emph{the data has to be fit for sharing}.
Data has to exhibit quality values, which fulfill a certain quality standard,
suitable for sharing it in a defined context.
Data consumers, especially in data markets, have thereby the possibility to
query data based on its quality. Hence, data can be rejected, if it does not
satisfy the negotiated standards.
In the motivating example, this could mean that the reading center rejects the
data, because important values are missing, i.e., the completeness of the data
set is too low.
Furthermore, data transparency plays a crucial role for enhancing trust for all
stakeholders.
\textbf{(DMR3)} \emph{Data transparency has to be enabled from the origin of the data until its usage.}
At any time in a data-driven pipeline, the current meaning of the data has to be
available, as well as metadata describing data transformations made by the
different components of the pipeline.
This explicitly includes traceability and transparency of algorithms and their
results (e.g., for data curation and integration, or for prediction).
Consider for example the use of data from cancer registries by researchers and
other registries. Both need to know explicitly how the data has been acquired
and modified to estimate its value for the research at hand.
Potential conflicts between data transparency and company secrets or privacy may
exist. Thus, transparency must be offered to all the stakeholders according to
their role in the DE, and in terms of consent management and usage control.
Hence, \textbf{(DMR4)} data quality management needs to \emph{take trade-offs
into account and provide dimensions and assessment metrics} that enable
stakeholders to rate the possible impact of, e.g., data curation.
Anonymized medical data for example may loose its value for further research if
important attributes are eliminated from a data set.
Lastly, data integration and querying over multiple data sources and across
organizations are required in a multitude of scenarios.
For this, mappings among data sources are defined either manually or
(semi-)automatically by schema matching.
For data quality management this implies several aspects, but basically
\textbf{(DMR5)} \emph{stakeholders should be part of the loop of data quality
assessment}.
They should be able to rate the quality of every step in a data-driven pipeline,
e.g., schema and entity mappings or query answers.
The automatic matching between huge medical taxonomies, e.g., for decision
support systems, may be very error-prone as the taxonomies per se have quality
problems.
\textbf{(DMR6)} \emph{The impact of adding a new component to a DE should be
measurable.} The DEs and their stakeholders should be able to rate the impact
of, e.g., the information gain of adding a new data set.
This is a crucial aspect especially when considering to pay for a costly data
set or when the integration of the data set requires a lot of upfront effort in
terms of data cleaning, transformation, or data integration.

\paragraph{Organizational-centric requirements}
In cases of sensitive data exchange and processing, data must be transparently
used according to organizations' policies, as well as its business models and
strategies.
\textbf{(OCR1)} \emph{Enabling data governance} is crucial for the appropriate
data exchange and sharing according to the organizations' strategies and 
business models.
\textbf{(OCR2)} \emph{Ensuring traceability of data sovereignty} is essential to
increase trust among stakeholders.
Again this has to be ensured throughout the whole data processing pipeline
including data quality assessment and curation.
For example, the willingness of patients to use applications or participate in
studies may be increased by giving them the opportunity to enforce access and
usage policies.
Furthermore, \textbf{(OCR3)} \emph{business, certification, and utility models
need to be established} to certify, based on data quality values and other
characteristics, the monetary value of exchanged and transformed data.
The monetary value of medical data is manifold, e.g., data from clinical studies
may be of interest to other parties, such as pharmaceutical or insurance
companies, to create or improve products.
\textbf{(OCR4)} \emph{Adherence to data and data processing standards} is
required to enhance DQ and interoperability across stakeholders, which is
specifically important in the medical domain. Standards such as
FHIR\footnote{\url{http://www.hl7.org/fhir/index.html}} have made an important
step forward reaching these goals in the clinical domain.
\textbf{(OCR5)} \emph{Flexible DQ management} for different coordination and
negotiation models among stakeholders (e.g., clinics, data scientists, and
insurance companies), and considering the evolution  of these models over time.

\paragraph{Legal and ethical requirements}
As shown in the motivating example, respecting personal data privacy and
security during data management, exchange, and analytics impairs requirements at
both, legal and ethical levels. Both categories of requirements are aligned with
the European Union guidelines for Trustworthy AI~\cite{EU2018}.
\textbf{(L\&ER1)}  \emph{Providing expressive legal frameworks for exchanged
data}, including legal references, responsibilities, licenses, and ethical
guidelines is essential. 
\textbf{(L\&ER2)} \emph{Accounting and mitigating bias and fairness} ensure that
the outcomes of the execution of the system components are independent of
sensitive attributes (e.g., gender, age, ethnicity, or health conditions) and
augment confidence in the impartiality of the system behavior.
\textbf{(L\&ER3)} \emph{Endeavouring safeness and robustness} of the decisions
made by each component that exchanges, processes, or analyzes data. Thus, the
deployment of data-driven pipelines and their outcomes will guarantee the
compliance with ethical guidelines (e.g., the respect of the patient consents),
and the misuse reduction that could conduce to data quality issues, data privacy
violation, and unfairness.
\textbf{(L\&ER4)} \emph{Enforcing data protection and ownership} safeguards
privacy, sovereignty, and legal compliance with licenses and regulations for
data sharing, exchange, and processing. Thus, the satisfaction of this
requirement will ensure that clinical data is used by the distinct parties as
indicated in the patient's consents.
\textbf{(L\&ER5)}  \emph{Pursuing diversity and non-discrimination} in data
collections shared, exchanged, and processed by data-driven pipelines. As a
result, the risk of excluding specific entities (e.g., patients with a given
health condition) is mitigated, and the chances of covering all the
representative entities of the population increase.
\textbf{(L\&ER6)} \emph{Trackability of regulations compliance} in the way that
each data-driven decision can be documented and validated in terms of legal
regulations, business models, and ethical guidelines.
Lastly, \textbf{(L\&ER7)} \emph{Trustworthiness and Reliability} of  data-driven
pipelines demand the accurate measurement, validation, and interpretation of
each of the decisions taken by the system components in compliance with legal
and ethical guidelines of the stakeholders. Thus, data owners (e.g., patients,
insurance companies, and researchers) will be able to trace management and
analysis methods performed over their data.

As illustrated in the next section, this paper positions \emph{networks of
knowledge-driven DEs} as alliance-driven decentralized infrastructures empowered
with components to satisfy the requirements listed above. Lastly,
section~\ref{sec:challenges} presents the challenges to be achieved to meet
these requirements.

\section{Knowledge-driven Data Ecosystems: Overview \& Architecture}
\label{sec:dataecosystems}

The literature defines DEs in different ways. For instance, Oliveira and Farias
L{\'{o}}scio~\cite{OliveiraL18} define Data Ecosystems as a ``set of networks
composed of autonomous actors, which consume, produce, or provide data or other
related resources.'' Other definitions add that the results created by the
consumption and processing of the data should return to the ecosystem
\cite{pollock2011building}. In \cite{bdva2020, otto2019}, the emergence of the
concept of DEs is traced and taxonomically situated among related concepts such
as Business Ecosystems, Digital Ecosystems, and Platform
Ecosystems~\cite{guggenberger2020ecosystem}.

Cappiello {\em et al.}~\cite{capiello_et_al:DR:2020:11845} synthesized the following
comprehensive definition of a data ecosystem \textit{DE} as a 4-tuple 
\textit{DE=$\langle$Data Sets, Data Operators, Meta-Data, Mappings$\rangle$}
where:
\begin{itemize}
    \item \textit{Data sets} can be structured or unstructured, can have
    different formats, e.g., CSV, JSON or tabular relations, and can be managed
    using different management systems.
    \item  \textit{Data Operators} are functions used for accessing or managing
    data in the data sets.
    \item \textit{Meta-Data} provides means for describing the DE context
    domain, can be used to specify the meaning of data and associated data
    operations.
    It comprises
    \begin{inparaenum}[\bf i\upshape)]
    \item \textit{Domain ontology}, providing a coherent and unified view of
    concepts, relationships, and constraints of the domain of knowledge, 
    associating formal semantics with the elements of the domain. If
    appropriate, several ontologies for different portions of the domain can be
    devised.
    \item \textit{Properties} that enable the definition of qualitative aspects
    for the elements of the ecosystem, such as quality and provenance
    requirements for data sets and operations.
    \item \textit{Descriptions} to associate annotations to the elements of the
    system for explaining relevant characteristics of data sets and operations.
    No specific formal language or vocabulary is required in descriptions.
    \end{inparaenum}
    \item \textit{Mappings} express correspondences among the different
    components of the data ecosystem. The mappings are as follows:
       \begin{inparaenum} [\bf i\upshape)]
    \item \textit{Mappings among ontologies} to represent associations among
    concepts in different ontologies constituting the domain ontology of the
    ecosystem.
        \item \textit{Mappings between data sets and ontology} to represent
        relations among the data in the DE data sets and the domain ontology, to
        allow for their interpretation in terms of the ontology.
    \end{inparaenum}
\end{itemize}

Data ecosystems can be further empowered with services that exploit the
knowledge encoded in the meta-data and operators to satisfy business
requirements, such as data transparency and traceability. We nam these
\emph{knowledge-driven data ecosystems}.
Services include query processing, data transformation, anonymization, data
quality assessment, or mapping generation. The following correspond to examples
of notable services:

\begin{itemize}
	\item \textit{Concept or mapping discovery}: identify a new concept or a new
	mapping using inductive reasoning and techniques from schema matching, taking
	into account aspects of uncertainty~\cite{gal2011uncertain}. Based on the
	result, the domain ontology and the mappings can be augmented.
	\item \textit{Data set curation}: identify the best way to keep humans in the
	loop in order to create a curated version of a data set in a DE
	(see~\cite{ackerman2019cognitive} for limitations of humans in matching).
	Services can also update the properties of a DE to indicate the provenance of
	the new curated data sets and manage new generated data from data
	transformation, analysis, and learning.
	\item \textit{Procedure synthesis}: construct new procedures out of data
	operators and other building blocks by composing existing services towards new
	goals. In complex and evolving systems, it is infeasible to program procedures
	and even queries without automatic support. Also, exploring repositories and
	libraries of existing procedures should be possible.
\end{itemize}
In our running example, stakeholders like clinics, insurance companies, and
researchers can create their own knowledge-driven DE. Each DE comprises data
sets and programs for accessing, managing, and analyzing their data.
Interoperability issues across data sets of a DE are solved in a unified view
represented in the DE ontology. Mappings between the data sets and the DE
ontology describe the meaning of the data sets. Moreover, the description of the
data operators enhances data transparency and provides the basis for tracking
down the computational methods executed against a DE.
\begin{figure}
	\centering
	\includegraphics[width=\textwidth]{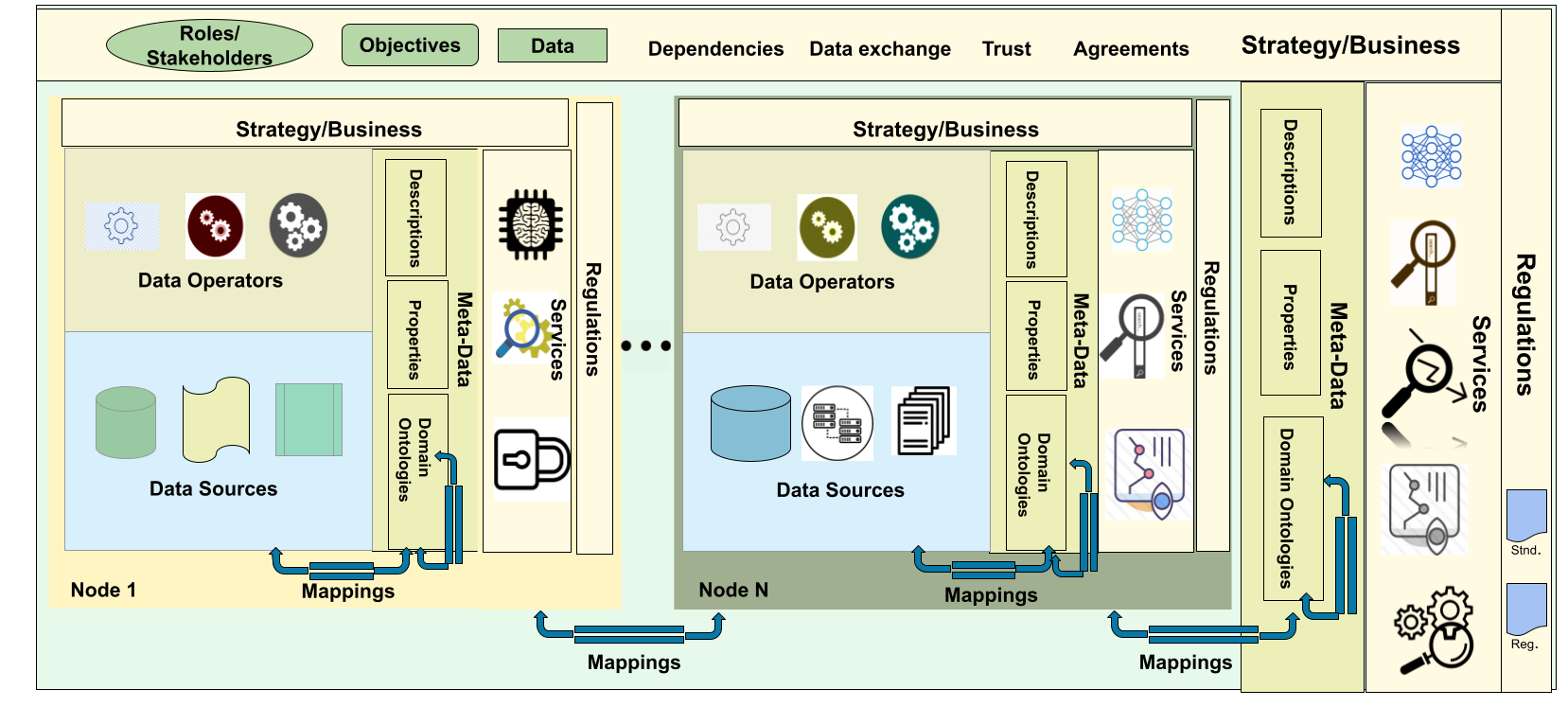}
	\caption{A Network of Data Ecosystems Empowered with Strategy and Business Models and Regulations.}
	\label{fig:DE}
\end{figure}

To enable collaboration across various knowledge-driven DEs and cope with
complex scenarios, a set of DEs can be connected in a network. For this, we
envision an \textit{ecosystem-wide} meta-data layer where the entire ecosystem
is described. \autoref{fig:DE} depicts a network where nodes and edges
correspond to knowledge-driven DEs and mappings among them, respectively. In
this configuration, the meta-data layer describes each of the nodes (i.e., DE)
in terms of descriptions, properties, and domain ontologies. The following
mappings can be defined among a network DEs:
\begin{itemize}
	\item \textit{Mappings between domain ontologies}: state correspondences among
	the domain ontologies of two nodes or between one node and the global meta-data
	layer.
	\item \textit{Mappings between properties}: describe relationships among
	properties in two nodes. For example, the provenance of two curated versions of
	a data set could be the same.
	\item \textit{Data set Mappings}: represent correspondences among data sets of
	different nodes.
\end{itemize}

Finally, knowledge-driven DEs can be enhanced with additional meta-data layers
to enable the description of business strategies and the access regulations.
\autoref{fig:DE} depicts the main components of a network of knowledge-driven
DEs empowered with these layers. As can be observed, this enriched version of a
network of DEs comprises:
\begin{inparaenum}
\item Ontological formalisms or causality models that enable the description of
the relationships between the data sets received as input and produced as output
for the services or operators of a DE.
\item  Meta-data describing business strategies that enable the definition of
the stakeholders of the network and their roles.
\item Objectives to be met and the dependencies among the tasks that need to be
performed to achieve these objectives.
\item Agreements for data exchange and criteria for trustworthiness.
\item Regulations and licenses for data access and for data privacy
preservation.
\item Services composing services of the nodes of the network.
\item Services able to monitor and explore  decisions taken by services and
operators.
\end{inparaenum}

A network of knowledge-driven DEs will facilitate controlled data exchange
across stakeholders in the running example. This network can be hosted and
maintained by the consortium of stakeholders. The meta-data layer specifies
alignments among the data sets in each DE. It enables the definition of business
models and access regulations to be satisfied when the data from one DE (e.g.,
the clinics DE) is transferred to other DEs (e.g., the DE of the insurance
companies or the researchers). Moreover, the formal descriptions of the data
sets and operators enhance transparency not only at an individual DE level
(e.g., clinics or insurance companies), but also in the network. Lastly,
services to monitor how exchanged data is processed in the various multi-site
clinical studies empower the network to verify if legal and ethical regulations
are fulfilled.
\section{Enabling Data Transparency in Data Ecosystems}
In this section, we discuss what are the challenges implementing the three
groups of requirements introduced in Section~\ref{sec:requirements} in the
network of knowledge-driven DEs.
In Section~\ref{sec:evaluation} we evaluate and rate how hard it is to meet the
requirements for the different types of DEs we defined in this section.

\subsection{Challenges of Enabling Data Transparency in Data Ecosystems}
\label{sec:challenges}

\textbf{Data Management Requirements.} 
Sharing heterogeneous data requires guarantees in terms of data quality and
transparency. As regards data quality, all the collaborative entities should
assess data quality using a common set of DQ services.

In fact, in order to get the maximum benefits from the participation to a DE,
actors should be able to identify, evaluate, and get the most suitable data for
the intended usage.
Starting from facilitating the access to available DE data, some existing
solutions propose that query processing over heterogeneous data sets rely on a
unified interface for overcoming interoperability issues, usually based on
meta-models \cite{Jeusfeld2010}. A few DEs have been proposed, mainly focusing
on data ingestion and meta-data extraction and management. Exemplary approaches
include Constance~\cite{HaiGQ16} and Ontario~\cite{EndrisRVA19}.
Data integration enables the transformation of heterogeneous data sources or
views under a unified access schema \cite{Lenzerini02}. Data integration systems
comprise data collection and curation steps and resort to record linkage, schema
matching, mapping, and data fusion to integrate data from a collection of data
sets ~\cite{CroceCLC20}. At the heart of the data integration realm lies the
matching task~\cite{BELLAHSENE2011}, in charge of aligning attributes of data
sources both at a schema and data level, in order to enable formal mappings.
Numerous algorithmic attempts ({\em matchers}) were suggested over the years for
efficient and effective integration ({\em
e.g.},~\cite{chen2018biggorilla,DO2002a,KENIG2013,konda2016magellan}).
Both practitioners and researchers also discussed data spaces as an appropriate
data integration concept for DEs. Data spaces do not require a common schema and
achieve data integration on a semantic level.

Moreover, only a small percentage of data integration systems provide causal
explanations to support traceability \cite{WangHM18}, as well as query
processing methods to navigate these explanations  \cite{RekatsinasRVZP19}
efficiently.
Existing rule-based approaches that allow for a declarative specification of
data transformation, integrity, and integration represent building-blocks for
tracking the validity of the domain constraints in all the data-driven pipeline
steps in a knowledge-driven DE. These approaches include rule-based entity
linking (e.g., \cite{SakorSPV20}), mapping-based tools to perform the process of
data integration (e.g., \cite{Jozashoori20}), and declarative languages like
SHACL \cite{Corman2019}, to describe integrity constraints.
Knowledge-driven DEs represent a new paradigm for data integration able to trace
and annotate provenance and causal relations existing during data ingestion,
curation, and integration \cite{HeyvaertMDV19}.

DQ challenges are also related to the data variety, and to the fact that the
description and measurement of data quality is highly subjective
\cite{Redman2001}, especially if data are used for completely different purposes
from the ones they were originally collected for (re-purposing). Algorithms able
to describe and assess the quality of very heterogeneous sources and very
different stakeholder views are required, together with an agreement about DQ
assessment standards in the DE.
These standards must include general DQ dimensions but also be derived from the
domain at hand to be accepted by the parties sharing the data.
The DQ assessment phase requires to provide meta-data and rules to support the
selection and reuse of data.
DQ should be also assessed on derived data. Challenges here are related to the
evaluation of the quality of the outputs of any transformation (e.g.,
aggregation, formulas, integration) and to the fact that in some cases quality
evaluation could be performed in a semi-automatic way.
Different stakeholders may have different views on them ({\em e.g.}, different
levels of granularity) and, as a consequence, their integration needs may vary.
The input offered by stakeholders regarding the quality of the integration is
therefore needed to tailor it to their needs.
Data transparency is enabled by a combination of accurate meta-data, including
provenance. The facilities to add these to a single, isolated DE are described
in Section \ref{sec:dataecosystems}.

Networks of DEs introduce further complications: i) meta-data descriptions
provided for similar entities by each DE may differ, and may require semantic
re-conciliation; ii) levels of transparency provided through provenance may
differ across the nodes.
A general challenge is therefore to achieve full transparency, or at least
formally characterize the boundaries of what is visible, in the presence of
``black spots'' in the global information flow across all nodes in the network.
Specific transparency challenges are also related to the difficulties in
tracking all the operations performed on data.
These new tasks call also for novel reasoning services, based on a sort of
reverse engineering process, which rewrites data source queries in terms of
global schema (or, ontology) expressions. Future technical support for
transparency, traceability, and usage policy enforcement could be developed from
distributed ledger technology (e.g., blockchains) and secure multi-party
computation (MPC) services.

\noindent
\textbf{Organization-centric Requirements.}
The enabling of data governance by several collaborating entities requires the
management of different governance models, different levels of data governance
maturity, and a clear attribution of responsibilities.
The traceability of data sovereignty requires process or dependency models which
define how data sovereignty has been executed, e.g., by usage control.
The challenge is here to define the content and granularity of information a
data owner needs to know based on her role according to regulations valid for
the domain.
A further challenge is to consider data quality assessment and curation, that
can add bias to the data and may also lead to unwanted usage of the data. We
need to  find business models and regulations which respect the interests of
involved stakeholders, adequately react to unwanted data usage, and increase the
trust between them.
A big challenge is how a data ecosystem can ensure that stakeholders comply to
data and data processing standards.
Usually, all stakeholders implement their own processes in multiple ways.
Enforcing certain standards in a data ecosystem or network of data ecosystems to
ensure data quality and data transparency across heterogeneous processes and
their interactions is challenging.
It requires to negotiate common standards and developing a model feasible for
SMEs and big stakeholders to implement.
Stakeholders could get a certification which testifies that they adhere to the
standards.
This makes the quality grade of data sources more transparent to data consumers.
As described by \citet{curry2018} data ecosystems may vary according to the
degree of interaction between stakeholders, coordination of data exchange, and
control over data sources. This variety heavily influences also data quality
management as it may substantially change the way to assess, monitor, and
improve data quality depending on the model.

\noindent
\textbf{Legal \& Ethical Requirements.}
The satisfaction of laws, fundamental rights, or ethical guidelines demands
traceability and certification of data-driven pipelines in a DE. The big
challenge is to devise services capable of certifying robustness in data
ecosystems according to the national and international legal norms for data
protection and fundamental rights, while safeguarding data sovereignty.
Furthermore,  formalisms, models, and computational algorithms able to
interoperate across various stakeholders represents grand challenges to ensure
data transparency.

\subsection{How can Data Ecosystems Fulfil Data Transparency Requirements?}
\label{sec:evaluation}
Concluding our analysis, we analyze the three DE architectures presented in
Section \ref{sec:dataecosystems} with respect to their potential in satisfying
the requirements and overcoming the challenges towards data transparency.
Each requirement is graded following a three-stars scheme:
 \begin{inparaenum}[\bf i\upshape)]
     \item $\star$ means that the requirement is unsatisfied;
     \item $\star$ $\star$ indicates that the data ecosystem has the potential
     to satisfy the requirement, but it is challenging; 
     \item $\star$ $\star$ $\star$ states that the data ecosystem has the
     potential to fully satisfy the requirement.
\end{inparaenum}
Each requirement is also evaluated in terms of three levels of satisfaction:
\begin{inparaenum}[\bf I\upshape)]
	\item \textbf{Complete:} The requirement is fully achieved.
	\item \textbf{Traceable:} The results of the requirement implementation can be
	traced down.
	\item \textbf{Verifiable:} The inspection, demonstration, test, and analysis of
	the requirement implementation can be verified.
\end{inparaenum}

\begin{table*}[ht!] 
\caption{\textbf{Three-stars Model for Requirement Satisfaction}.}
\label{DEAnalysis}
\begin{adjustbox}{width=\textwidth}
\begin{tabular}{l|c|c|c|c|c|c|c|c|c|}
\cline{2-10} & \multicolumn{9}{c|}{\cellcolor[HTML]{7C98AA}\textbf{Data Management Requirements}} \\ \cline{2-10} 
& \multicolumn{3}{c|}{\cellcolor[HTML]{AED6F1} \textbf{Data Ecosystem}}  & \multicolumn{3}{c|}{\cellcolor[HTML]{AED6F1}\textbf{Knowledge-Driven Data Ecosystem}} &\multicolumn{3}{c|}{\cellcolor[HTML]{AED6F1}\textbf{Network of Knowledge-Driven Data Ecosystems}} \\ \cline{2-10}
& \multicolumn{1}{l|}{\cellcolor[HTML]{68CBD0} \textbf{Complete}}   & \multicolumn{1}{l|}{\cellcolor[HTML]{68CBD0} \textbf{Traceable}}   & \multicolumn{1}{c|}{\cellcolor[HTML]{68CBD0} \textbf{Verifiable}}   & \multicolumn{1}{l|}{\cellcolor[HTML]{DAE8FC}\textbf{Complete}}   & \multicolumn{1}{l|}{\cellcolor[HTML]{DAE8FC}\textbf{Traceable}}   & \multicolumn{1}{c|}{\cellcolor[HTML]{DAE8FC}\textbf{Verifiable}}   & \multicolumn{1}{l|}{\cellcolor[HTML]{CBCEFB}\textbf{Complete}}   & \multicolumn{1}{l|}{\cellcolor[HTML]{CBCEFB}\textbf{Traceable}}   & \multicolumn{1}{c|}{\cellcolor[HTML]{CBCEFB}\textbf{Verifiable}}      \\ \hline
\multicolumn{1}{|l|}{\cellcolor[HTML]{FFFFC7}DMR1}                      & \multicolumn{1}{c|}{\cellcolor[HTML]{68CBD0} $\star$ $\star$ $\star$ } & \multicolumn{1}{c|}{\cellcolor[HTML]{68CBD0} $\star$  } & \multicolumn{1}{c|}{\cellcolor[HTML]{68CBD0} $\star$  } & \multicolumn{1}{c|}{\cellcolor[HTML]{DAE8FC} $\star$ $\star$ $\star$} & \multicolumn{1}{c|}{\cellcolor[HTML]{DAE8FC} $\star$ $\star$ } & \multicolumn{1}{c|}{\cellcolor[HTML]{DAE8FC} $\star$ $\star$ } & \multicolumn{1}{c|}{\cellcolor[HTML]{CBCEFB} $\star$ $\star$ $\star$ } & \multicolumn{1}{c|}{\cellcolor[HTML]{CBCEFB} $\star$ $\star$ } & \multicolumn{1}{c|}{\cellcolor[HTML]{CBCEFB} $\star$ $\star$}  \\ \hline
\multicolumn{1}{|l|}{\cellcolor[HTML]{FFFFC7}DMR2} & \cellcolor[HTML]{68CBD0}$\star$ $\star$ $\star$ & \cellcolor[HTML]{68CBD0}$\star$ 
& \cellcolor[HTML]{68CBD0}$\star$ 
& \cellcolor[HTML]{DAE8FC}$\star$ $\star$ $\star$   & \cellcolor[HTML]{DAE8FC}$\star$ $\star$  & \cellcolor[HTML]{DAE8FC}$\star$ $\star$ 
& \cellcolor[HTML]{CBCEFB} $\star$ $\star$ $\star$  & \cellcolor[HTML]{CBCEFB}$\star$ $\star$ & \cellcolor[HTML]{CBCEFB}$\star$ $\star$  \\ \hline
\multicolumn{1}{|l|}{\cellcolor[HTML]{FFFFC7}DMR3} & \cellcolor[HTML]{68CBD0}$\star$ $\star$ $\star$  & \cellcolor[HTML]{68CBD0}$\star$ 
& \cellcolor[HTML]{68CBD0}$\star$ 
& \cellcolor[HTML]{DAE8FC}$\star$ $\star$ $\star$   & \cellcolor[HTML]{DAE8FC}$\star$ $\star$    & \cellcolor[HTML]{DAE8FC}$\star$ $\star$   & \cellcolor[HTML]{CBCEFB}$\star$ $\star$ $\star$  & \cellcolor[HTML]{CBCEFB}$\star$ $\star$   & \cellcolor[HTML]{CBCEFB}$\star$ $\star$   \\ \hline
\multicolumn{1}{|l|}{\cellcolor[HTML]{FFFFC7}DMR4} & \cellcolor[HTML]{68CBD0}$\star$ $\star$ $\star$  & \cellcolor[HTML]{68CBD0}$\star$ 
& \cellcolor[HTML]{68CBD0}$\star$ 
& \cellcolor[HTML]{DAE8FC}$\star$ $\star$ $\star$   & \cellcolor[HTML]{DAE8FC}$\star$ $\star$    & \cellcolor[HTML]{DAE8FC}$\star$ $\star$
& \cellcolor[HTML]{CBCEFB}$\star$ $\star$ $\star$  & \cellcolor[HTML]{CBCEFB}$\star$ $\star$   & \cellcolor[HTML]{CBCEFB}$\star$ $\star$  \\ \hline
\multicolumn{1}{|l|}{\cellcolor[HTML]{FFFFC7}DMR5} & \cellcolor[HTML]{68CBD0}$\star$ $\star$   & \cellcolor[HTML]{68CBD0}$\star$ 
& \cellcolor[HTML]{68CBD0}$\star$ 
& \cellcolor[HTML]{DAE8FC}$\star$ $\star$ $\star$   & \cellcolor[HTML]{DAE8FC}$\star$ $\star$    & \cellcolor[HTML]{DAE8FC}$\star$ $\star$
& \cellcolor[HTML]{CBCEFB}$\star$ $\star$ $\star$  & \cellcolor[HTML]{CBCEFB}$\star$ $\star$   & \cellcolor[HTML]{CBCEFB}$\star$ $\star$  \\ \hline
\multicolumn{1}{|l|}{\cellcolor[HTML]{FFFFC7}DMR6} & \cellcolor[HTML]{68CBD0}$\star$ $\star$
& \cellcolor[HTML]{68CBD0}$\star$ 
& \cellcolor[HTML]{68CBD0}$\star$ 
& \cellcolor[HTML]{DAE8FC}$\star$ $\star$ $\star$   & \cellcolor[HTML]{DAE8FC}$\star$ $\star$   & \cellcolor[HTML]{DAE8FC}$\star$ $\star$
& \cellcolor[HTML]{CBCEFB}$\star$ $\star$ $\star$  & \cellcolor[HTML]{CBCEFB}$\star$ $\star$   & \cellcolor[HTML]{CBCEFB}$\star$ $\star$  \\ \hline
\cline{2-10} & \multicolumn{9}{c|}{\cellcolor[HTML]{7C98AA}\textbf{Organizational Challenges}} \\ \cline{2-10} 
& \multicolumn{3}{c|}{\cellcolor[HTML]{AED6F1} \textbf{Data Ecosystem}}  & \multicolumn{3}{c|}{\cellcolor[HTML]{AED6F1}\textbf{Knowledge-Driven Data Ecosystem}} &\multicolumn{3}{c|}{\cellcolor[HTML]{AED6F1}\textbf{Network of Knowledge-Driven Data Ecosystems}} \\ \cline{2-10} 
& \multicolumn{1}{l|}{\cellcolor[HTML]{68CBD0}\textbf{Complete}}   & \multicolumn{1}{l|}{\cellcolor[HTML]{68CBD0}\textbf{Traceable}}   & \multicolumn{1}{c|}{\cellcolor[HTML]{68CBD0}\textbf{Verifiable}}   & \multicolumn{1}{l|}{\cellcolor[HTML]{DAE8FC}\textbf{Complete}}   & \multicolumn{1}{l|}{\cellcolor[HTML]{DAE8FC}\textbf{Traceable}}   & \multicolumn{1}{c|}{\cellcolor[HTML]{DAE8FC}\textbf{Verifiable}}   & \multicolumn{1}{l|}{\cellcolor[HTML]{CBCEFB}\textbf{Complete}}   & \multicolumn{1}{l|}{\cellcolor[HTML]{CBCEFB}\textbf{Traceable}}   & \multicolumn{1}{c|}{\cellcolor[HTML]{CBCEFB}\textbf{Verifiable}}      \\ \hline
\multicolumn{1}{|l|}{\cellcolor[HTML]{FFFFC7}OCR1}                      & \multicolumn{1}{c|}{\cellcolor[HTML]{68CBD0}  $\star$  } & \multicolumn{1}{c|}{\cellcolor[HTML]{68CBD0} $\star$ } & \multicolumn{1}{c|}{\cellcolor[HTML]{68CBD0} $\star$} & \multicolumn{1}{c|}{\cellcolor[HTML]{DAE8FC} $\star$  } & \multicolumn{1}{c|}{\cellcolor[HTML]{DAE8FC} $\star$ } & \multicolumn{1}{c|}{\cellcolor[HTML]{DAE8FC} $\star$} & \multicolumn{1}{c|}{\cellcolor[HTML]{CBCEFB} $\star$ $\star$ $\star$} & \multicolumn{1}{c|}{\cellcolor[HTML]{CBCEFB} $\star$ $\star$ } & \multicolumn{1}{c|}{\cellcolor[HTML]{CBCEFB} $\star$ $\star$ }  \\ \hline
\multicolumn{1}{|l|}{\cellcolor[HTML]{FFFFC7}OCR2} & \cellcolor[HTML]{68CBD0}$\star$  & \cellcolor[HTML]{68CBD0}$\star$  & \cellcolor[HTML]{68CBD0}$\star$ 
& \cellcolor[HTML]{DAE8FC}$\star$  & \cellcolor[HTML]{DAE8FC}$\star$ 
& \cellcolor[HTML]{DAE8FC}$\star$ 
& \cellcolor[HTML]{CBCEFB}$\star$ $\star$ $\star$  & \cellcolor[HTML]{CBCEFB}$\star$ $\star$
& \cellcolor[HTML]{CBCEFB}$\star$ $\star$
\\ \hline
\multicolumn{1}{|l|}{\cellcolor[HTML]{FFFFC7}OCR3} & \cellcolor[HTML]{68CBD0} $\star$ & \cellcolor[HTML]{68CBD0}$\star$  & \cellcolor[HTML]{68CBD0}$\star$  & \cellcolor[HTML]{DAE8FC}$\star$  & \cellcolor[HTML]{DAE8FC}$\star$   & \cellcolor[HTML]{DAE8FC}$\star$ 
& \cellcolor[HTML]{CBCEFB}$\star$ $\star$ $\star$  & \cellcolor[HTML]{CBCEFB}$\star$ $\star$ 
& \cellcolor[HTML]{CBCEFB}$\star$ $\star$  \\ \hline
\multicolumn{1}{|l|}{\cellcolor[HTML]{FFFFC7}OCR4} & \cellcolor[HTML]{68CBD0}$\star$ $\star$ $\star$& \cellcolor[HTML]{68CBD0}$\star$  & \cellcolor[HTML]{68CBD0}$\star$  & \cellcolor[HTML]{DAE8FC}$\star$ $\star$ $\star$
& \cellcolor[HTML]{DAE8FC}$\star$ $\star$    & \cellcolor[HTML]{DAE8FC}$\star$ $\star$ 
& \cellcolor[HTML]{CBCEFB}$\star$ $\star$ $\star$
& \cellcolor[HTML]{CBCEFB}$\star$ $\star$ & \cellcolor[HTML]{CBCEFB}$\star$ $\star$   \\ \hline
\multicolumn{1}{|l|}{\cellcolor[HTML]{FFFFC7}OCR5} & \cellcolor[HTML]{68CBD0}$\star$ & \cellcolor[HTML]{68CBD0}$\star$   & \cellcolor[HTML]{68CBD0}$\star$ 
& \cellcolor[HTML]{DAE8FC}$\star$ & \cellcolor[HTML]{DAE8FC}$\star$  & \cellcolor[HTML]{DAE8FC}$\star$ 
& \cellcolor[HTML]{CBCEFB}$\star$ $\star$ $\star$  & \cellcolor[HTML]{CBCEFB}$\star$ $\star$   & \cellcolor[HTML]{CBCEFB}$\star$ $\star$  \\ \hline
\cline{2-10} & \multicolumn{9}{c|}{\cellcolor[HTML]{7C98AA}\textbf{Legal \& Ethical Challenges}} \\ \cline{2-10} 
& \multicolumn{3}{c|}{\cellcolor[HTML]{AED6F1} \textbf{Data Ecosystem}}  & \multicolumn{3}{c|}{\cellcolor[HTML]{AED6F1}\textbf{Knowledge-Driven Data Ecosystem}} &\multicolumn{3}{c|}{\cellcolor[HTML]{AED6F1}\textbf{Network of Knowledge-Driven Data Ecosystems}} \\ \cline{2-10} 
& \multicolumn{1}{l|}{\cellcolor[HTML]{68CBD0}\textbf{Complete}}   & \multicolumn{1}{l|}{\cellcolor[HTML]{68CBD0}\textbf{Traceable}}   & \multicolumn{1}{c|}{\cellcolor[HTML]{68CBD0}\textbf{Verifiable}}   & \multicolumn{1}{l|}{\cellcolor[HTML]{DAE8FC}\textbf{Complete}}   & \multicolumn{1}{l|}{\cellcolor[HTML]{DAE8FC}\textbf{Traceable}}   & \multicolumn{1}{c|}{\cellcolor[HTML]{DAE8FC}\textbf{Verifiable}}   & \multicolumn{1}{l|}{\cellcolor[HTML]{CBCEFB}\textbf{Complete}}   & \multicolumn{1}{l|}{\cellcolor[HTML]{CBCEFB}\textbf{Traceable}}   & \multicolumn{1}{c|}{\cellcolor[HTML]{CBCEFB}\textbf{Verifiable}}      \\ \hline
\multicolumn{1}{|l|}{\cellcolor[HTML]{FFFFC7}L\&ER1}                      & \multicolumn{1}{c|}{\cellcolor[HTML]{68CBD0}  $\star$  } & \multicolumn{1}{c|}{\cellcolor[HTML]{68CBD0} $\star$ } & \multicolumn{1}{c|}{\cellcolor[HTML]{68CBD0} $\star$} & \multicolumn{1}{c|}{\cellcolor[HTML]{DAE8FC} $\star$  } & \multicolumn{1}{c|}{\cellcolor[HTML]{DAE8FC} $\star$ } & \multicolumn{1}{c|}{\cellcolor[HTML]{DAE8FC} $\star$  } & \multicolumn{1}{c|}{\cellcolor[HTML]{CBCEFB} $\star$ $\star$ $\star$} & \multicolumn{1}{c|}{\cellcolor[HTML]{CBCEFB} $\star$ $\star$ } & \multicolumn{1}{c|}{\cellcolor[HTML]{CBCEFB} $\star$ $\star$ }  \\ \hline
\multicolumn{1}{|c|}{\cellcolor[HTML]{FFFFC7}L\&ER2} & \multicolumn{1}{c|}{\cellcolor[HTML]{68CBD0}$\star$ $\star$ } & \multicolumn{1}{c|}{\cellcolor[HTML]{68CBD0}$\star$} 
& \multicolumn{1}{c|}{\cellcolor[HTML]{68CBD0}$\star$} 
& \multicolumn{1}{c|}{\cellcolor[HTML]{DAE8FC}$\star$ $\star$ } & \multicolumn{1}{c|}{\cellcolor[HTML]{DAE8FC}$\star$ $\star$ }  & \multicolumn{1}{c|}{\cellcolor[HTML]{DAE8FC}$\star$ $\star$ }  & 
\multicolumn{1}{c|}{\cellcolor[HTML]{CBCEFB}$\star$ $\star$  } & \multicolumn{1}{c|}{\cellcolor[HTML]{CBCEFB}$\star$ $\star$  }& 
\multicolumn{1}{c|}{\cellcolor[HTML]{CBCEFB}$\star$ $\star$ } \\ \hline
\multicolumn{1}{|c|}{\cellcolor[HTML]{FFFFC7}L\&ER3} & \multicolumn{1}{c|}{\cellcolor[HTML]{68CBD0}$\star$ }
& \multicolumn{1}{c|}{\cellcolor[HTML]{68CBD0}$\star$ }
& \multicolumn{1}{c|}{\cellcolor[HTML]{68CBD0}$\star$ }
& \multicolumn{1}{c|}{\cellcolor[HTML]{DAE8FC}$\star$ }  & \multicolumn{1}{c|}{\cellcolor[HTML]{DAE8FC}$\star$} 
& \multicolumn{1}{c|}{\cellcolor[HTML]{DAE8FC}$\star$ }
& \multicolumn{1}{c|}{\cellcolor[HTML]{CBCEFB}$\star$ $\star$}& \multicolumn{1}{c|}{\cellcolor[HTML]{CBCEFB}$\star$ $\star$}   & \multicolumn{1}{c|}{\cellcolor[HTML]{CBCEFB}$\star$ $\star$ }\\ \hline
\multicolumn{1}{|c|}{\cellcolor[HTML]{FFFFC7}L\&ER4} & \multicolumn{1}{c|}{\cellcolor[HTML]{68CBD0}$\star$ }
& \multicolumn{1}{c|}{\cellcolor[HTML]{68CBD0}$\star$ }
& \multicolumn{1}{c|}{\cellcolor[HTML]{68CBD0}$\star$}
& \multicolumn{1}{c|}{\cellcolor[HTML]{DAE8FC}$\star$ }  & \multicolumn{1}{c|}{\cellcolor[HTML]{DAE8FC}$\star$ }  & \multicolumn{1}{c|}{\cellcolor[HTML]{DAE8FC}$\star$ }  & \multicolumn{1}{c|}{\cellcolor[HTML]{CBCEFB}$\star$ $\star$ }  & \multicolumn{1}{c|}{\cellcolor[HTML]{CBCEFB}$\star$ $\star$}  & \multicolumn{1}{c|}{\cellcolor[HTML]{CBCEFB}$\star$ $\star$ } 
\\ \hline
\multicolumn{1}{|c|}{\cellcolor[HTML]{FFFFC7}L\&ER5} & \multicolumn{1}{c|}{\cellcolor[HTML]{68CBD0}$\star$ $\star$ } & \multicolumn{1}{c|}{\cellcolor[HTML]{68CBD0}$\star$ }
& \multicolumn{1}{c|}{\cellcolor[HTML]{68CBD0}$\star$ }
& \multicolumn{1}{c|}{\cellcolor[HTML]{DAE8FC}$\star$ $\star$  } & \multicolumn{1}{c|}{\cellcolor[HTML]{DAE8FC}$\star$ $\star$}  
& \multicolumn{1}{c|}{\cellcolor[HTML]{DAE8FC}$\star$ $\star$}
& \multicolumn{1}{c|}{\cellcolor[HTML]{CBCEFB}$\star$ $\star$ } & \multicolumn{1}{c|}{\cellcolor[HTML]{CBCEFB}$\star$ $\star$  }
& \multicolumn{1}{c|}{\cellcolor[HTML]{CBCEFB}$\star$ $\star$ } \\
\hline
\multicolumn{1}{|c|}{\cellcolor[HTML]{FFFFC7}L\&ER6} & \cellcolor[HTML]{68CBD0}$\star$ 
& \multicolumn{1}{c|}{\cellcolor[HTML]{68CBD0}$\star$} 
& \multicolumn{1}{c|}{\cellcolor[HTML]{68CBD0}$\star$ }
& \multicolumn{1}{c|}{\cellcolor[HTML]{DAE8FC}$\star$ }
& \multicolumn{1}{c|}{\cellcolor[HTML]{DAE8FC}$\star$ }
& \multicolumn{1}{c|}{\cellcolor[HTML]{DAE8FC}$\star$}
& \multicolumn{1}{c|}{\cellcolor[HTML]{CBCEFB}$\star$ $\star$ } & \multicolumn{1}{c|}{\cellcolor[HTML]{CBCEFB}$\star$ $\star$  }& \multicolumn{1}{c|}{\cellcolor[HTML]{CBCEFB}$\star$ $\star$ } \\
\hline
\multicolumn{1}{|c|}{\cellcolor[HTML]{FFFFC7}L\&ER7} & \multicolumn{1}{c|}{\cellcolor[HTML]{68CBD0}$\star$ }
& \multicolumn{1}{c|}{\cellcolor[HTML]{68CBD0}$\star$ }
& \multicolumn{1}{c|}{\cellcolor[HTML]{68CBD0}$\star$}
& \multicolumn{1}{c|}{\cellcolor[HTML]{DAE8FC}$\star$ }
& \multicolumn{1}{c|}{\cellcolor[HTML]{DAE8FC}$\star$ }
& \multicolumn{1}{c|}{\cellcolor[HTML]{DAE8FC}$\star$} 
& \multicolumn{1}{c|}{\cellcolor[HTML]{CBCEFB}$\star$ $\star$}
& \multicolumn{1}{c|}{\cellcolor[HTML]{CBCEFB}$\star$ $\star$} & \multicolumn{1}{c|}{\cellcolor[HTML]{CBCEFB}$\star$ $\star$ } \\
\hline
\end{tabular}
\end{adjustbox}
\label{tab:performance_teets}
\end{table*}

\autoref{DEAnalysis} summarizes the analysis of the types of DEs in Section
\ref{sec:dataecosystems}.
We can observe that the baseline architecture of DEs has the potential to
fulfill several requirements (i.e., DMR1-DMR6, OCR4, L\&ECR2, and L\&ECR5).
However, since these DEs are only equipped with data sets, operators, and
meta-data, it is challenging for them to keep the stakeholders in the loop
during the data quality rating or for assessing the impact of adding new
components. For the same reason, these DEs cannot trace or validate the
satisfaction of none of the requirements.
In contrast, the other two types of DEs are able to fully satisfy the data
management requirements (i.e., DMR1-DMR6). Nevertheless, requirement
traceability and verifiability still remain a challenge because of the multiple
problems of interoperability, data access, and legal regulations imposed by the
stakeholders of a knowledge-driven DE or each individual node in a network of
DEs. Moreover, a single knowledge-driven DE cannot interact with other DE or
circulate their business, regulations, or strategies. As a result, most
organization-centric and legal and ethical requirements cannot be satisfied, or
if so, it is very challenging. Lastly, networks of knowledge-driven DEs are
equipped with meta-data, services, and strategic and business models that
facilitate the description of each node and the documentation of the
negotiations required to exchange across the network. Thus, despite traceability
and verifiability are challenging, these DEs are the only ones furnished with
components to enable data transparency.

We hope that this analysis contributes to the understanding of data transparency
challenges. We also aim to encourage the research community to develop trustable
networks of knowledge-driven DEs, enabling, thus, DQ management, data
governance, and sovereignty, as well as mechanisms to trace and verify the
requirement fulfillment.
 
\section{Conclusion}
\label{sec:conclusion}

In this work, we have tackled the challenges that DEs face on their way to
become ``smarter," equipped with a knowledge layer. In particular, we focused on
data quality and data transparency challenges. Using the motivating example of
multi-site clinical studies, we have outlined six data management requirements,
five organizational-centric requirements, and seven legal and ethical
requirements. We then presented a specific architecture from which data quality
challenges were derived and discussed. Table~\ref{tab:performance_teets}
summarizes the discussion by presenting for each of three types of DEs to what
extent each of the requirements can be completed, traced, and verified.

With the increasing need for integrated data sets and infrastructures to support
DEs, we expect their impact on organizations to increase. As data quality in
general and data transparency in particular, become a significant issue in data
management, we hope this work offers a guideline for researchers and
practitioners when investigating developments of knowledge-driven DEs.
\section*{Acknowledgements}
The authors are grateful to the Dagstuhl team for hosting us in September 2019
(Dagstuhl Seminar 19391). Initial ideas that serve as a basis for this paper
were originated and discussed there.
Gal also acknowledges the support of the Benjamin and Florence Free Chair.
Lenzerini acknowledges the support of MUR-PRIN project ``HOPE'', grant n.
2017MMJJRE, and of EU under the H2020-EU.2.1.1 project TAILOR, grant id. 952215.
Vidal acknowledges the support of the EU H2020 project iASiS, grant id. 727658
and CLARIFY grant id. 875160. Geisler acknowledges the support of the German
Innovation Fund project SALUS, grant id. 01NVF18002. This work has also been
supported by the German Federal Ministry of Education and Research (BMBF) in the
context of the InDaSpacePlus project (grant id. 01IS17031), Fraunhofer Cluster
of Excellence "Cognitive Internet Technologies" (CCIT) and by the Deutsche
Forschungsgemeinschaft (DFG) under Germany's Excellence Strategy - EXC-2023
Internet of Production - 390621612. Pernici acknowledges the support of the EU
H2020 Crowd4SDG project, grant id 872944.

\bibliographystyle{plainnat}
\bibliography{Biblio}
\end{document}